\documentclass[aps,twocolumn,showpacs]{revtex4}
\usepackage{dcolumn}
\usepackage{graphicx}
\usepackage{amsmath}
\usepackage{amsfonts}
\usepackage{amssymb}
\usepackage{psfrag}
\usepackage{wrapfig}
\usepackage{subfigure}
\usepackage{makeidx}
\usepackage{bm}
\usepackage{epsf}
\usepackage{multirow}
\DeclareMathOperator{\sech}{sech}

\begin{document}

\title{Solitons in Nonlinear Systems and Eigen-states in Quantum Wells}
\author{Li-Chen Zhao$^{1,2}$}
\author{Zhan-Ying Yang$^{1,2}$}
\author{Wen-Li Yang$^{1,2,3}$}
\address{$^{1}$School of Physics, Northwest University, Xi'an, 710069, China}
\address{$^{2}$Shaanxi Key Laboratory for Theoretical Physics Frontiers, Xi'an, 710069, China}
\address{$^{3}$Institute of Modern Physics, Northwest University, Xi¡¯an 710069, China}
\date{\today}
\begin{abstract}

We study the relations between solitons of nonlinear Schr\"{o}dinger equation described systems and eigen-states of linear Schr\"{o}dinger equation with some quantum wells.  Many different non-degenerated solitons are re-derived from the eigen-states in the quantum wells. We show that the vector solitons for coupled system with attractive interactions correspond to the identical eigen-states with the ones of coupled systems with repulsive interactions. The energy eigenvalues of them seem to be different, but they can be reduced to identical ones in the same quantum wells.  The non-degenerated solitons for multi-component systems can be used to construct much abundant degenerated solitons in more components coupled cases. On the other hand, we demonstrate soliton solutions in nonlinear systems can be also used to solve the eigen-problems of quantum wells. As an example, we present eigenvalue and eigen-state in a complicated quantum well for which the Hamiltonian belongs to  the non-Hermitian Hamiltonian having Parity-Time symmetry. We further present the ground state and the first exited state in an asymmetric quantum double-well from asymmetric solitons. Based on these results, we expect that many nonlinear physical systems can be used to observe the quantum states evolution of quantum wells, such as water wave tank, nonlinear fiber, Bose-Einstein condensate, and even plasma, although some of them are classical physical systems.  These relations provide another different way to understand the stability of solitons in nonlinear Schr\"{o}dinger equation described systems, in contrast to the balance between dispersion and nonlinearity.

\end{abstract}
\maketitle

\section{Introduction}
Solitons have been found in many different physical systems \cite{Zabusky,Kuznetsov,Agrawal,Osborne,Kevrekidis}, such as bright soliton, dark soliton, breather-like solitons, and abundant vector solitons.
The prototypical model for the study of these solitons is the
nonlinear Schr\"{o}dinger (NLS) equation with attractive or repulsive interactions. Many experiments suggested that the NLS model described well the evolution dynamics of many different nonlinear systems, such as water wave tank \cite{water}, optical fiber\cite{fiber}, Bose-Einstein condensate\cite{BEC}, etc. It is noted that the nonlinear terms can be understood as some certain quantum potentials \cite{Nogami,NA,Zhaot}. If the density profiles depend on time, the NLS evolution corresponds to LS with time-dependent quantum wells. It is also hard to solve LS with time-dependent quantum wells, therefore we focus the eigen-problems of LS with time-independent potentials.   If the density profile of nonlinear term is time-independent, then the potentials will  become also time-independent, and the NLS can be related with the eigen-problems of LS with the quantum potentials \cite{Nogami,NA}.  This provides possibilities to establish a correspondence relations between soliton states and the eigen-states in quantum wells.  The relations between them are nontrivial to be established, since they enable us to obtain more abundant dynamics of localized waves in nonlinear systems \cite{Mateo}, and solve the eigen-problems in more complicated quantum wells. They also provide possibilities to investigate the evolution of quantum states in classical nonlinear systems.

In this paper, we study the quantitative relations between solitons of NLS equation with attractive or repulsive interactions and eigen-states of LS equation with quantum wells. The discussions are made on static solions for simplicity. It should be noted that soliton with velocities can be related with static soliton through Galilean transformation, since the NLS equations admit Galilean symmetry. Therefore, the relations also hold well for solitons with velocities, just with the Galilean transformation holds. We start from the simplest scalar NLS with attractive interaction to establish the relations, and then generalize it to multi-component NLS  cases with attractive interactions. Especially, we show that vector solitons in multi-component NLS with repulsive interaction can be generated from the eigen-problem in  identical quantum wells for NLS with attractive interactions. We establish the explicit relations up to general $N$-component coupled case, and the results are summarized in the Table I.  The non-degenerated soliton for $M$-component systems can be used to construct abundant degenerated soliton in more than $M$ components coupled cases. On the other hand, we also demonstrate soliton solution  in nonlinear systems can be  used to solve the eigen-problems of quantum wells. This paves the ways to solve the LS with complicated potentials with the aid of well-developed techniques for solving NLS equations.

The paper are organized as follows. In Sec. II, we describe the simple idea for studying the relations between soliton of NLS and eigen-states of LS with certain quantum wells. In Sec. III, we establish the relations between non-degenerated soliton of vector NLS with attractive interactions and the eigen-states in quantum wells. Many different soliton presented previously are re-derived in much simpler way from the eigen-states in quantum wells. The eigenvalues of solitons are clarified clearly. In Sec. IV, we derive non-degenerated soliton solution of vector NLS with repulsive interactions from the identical quantum wells with the vector NLS with attractive interactions. They admit identical eigenvalues and eigen-states with the ones for attractive interactions cases. Then, we discuss how degenerated soliton for $N$-component coupled NLS with attractive or repulsive interactions can be generated from the non-degenerated solitons of $M$-component coupled NLS, with $N>M$, in Sec. V.   On the other hand, we  present eigenvalue and its eigen-state in two complicated quantum wells form the soliton solution of NLS in Sec. VI.  This paves the ways to solve the LS with complicated potentials with the aid of well-developed techniques for solving NLS equations. Finally, we present our conclusion and discussions in Sec. VII.

\begin{table*}[htp]
\large
\centering
\renewcommand{\arraystretch}{1}
\begin{tabular}{|c|c|c|c|}
\hline
The quantum well of LS  & Eigenvalues\  &\ \ N-c. coupled NLS with AI\ \ & \ \ N-c. coupled NLS with RI\ \ \\
\hline
\multirow{9}{*}{$-\frac{N(N-1)}{2}\  f\ \sech^2[\sqrt{f}\  x]$} &$ 0$& \multicolumn{2}{c|}{$(N-1)$-valley dark soliton} \\
  &$ -f/2$&\multicolumn{2}{c|}{$(N-1)$-hump bright soliton} \\
  & $-2\ f$&\multicolumn{2}{c|}{$(N-2)$-hump bright soliton} \\
  & $-9\ f/2$&\multicolumn{2}{c|}{$(N-3)$-hump bright soliton} \\
  & . &\multicolumn{2}{c|}{.} \\
  & . &\multicolumn{2}{c|}{.} \\
  & . &\multicolumn{2}{c|}{.} \\
  & $-(N-2)^2f/2$ &\multicolumn{2}{c|}{double-hump bright soliton} \\
  & $-(N-1)^2f/2$ &\multicolumn{2}{c|}{single-hump bright soliton} \\
\hline
\end{tabular}
\caption{The correspondence relations between  the eigenvalues of linear Sch\"{o}dinger (LS) equation with a quantum well and non-degenerated solitons of $N$-component coupled nonlinear Sch\"{o}dinger (NLS) equation with attractive or repulsive interactions. The correspondence relations enable us to obtain many many different types of solitons in NLS described systems, and\emph{ provide a unified way to derive soliton solutions for both attractive and repulsive interaction cases.} It should be noted that the coefficients of these soliton solution are different for the attractive case and repulsive case.  Especially, \emph{the non-degenerated soliton can be further used to generate degenerated solitons and beating solions for more than $N$ components coupled cases.} The ``N-c.", ``AI" and ``RI" denote N-component, attractive interactions, and repulsive interactions respectively in the table. }
\end{table*}

\section{A simple idea for identifying the relations between solitons and eigen-states in quantum wells}
With dimensionless unit, the simplest NLS   equation can be written as follows,
\begin{equation}
i\frac{\partial \psi(x,t)}{\partial t}+\frac{1}{2} \frac{\partial^2
\psi(x,t)}{\partial x^2}+\gamma |\psi(x,t)|^2 \psi(x,t) =0,
\end{equation}
which has been studied widely in soliton fields \cite{Kuznetsov,Agrawal,Osborne,Kevrekidis}. If $\gamma=0$, the equation will become a LS equation with no potentials. It is not possible to obtain bound state in this case. If $\gamma=1$, the NLS will admit bright soliton which is a hump density on zero background; if $\gamma=-1$, the NLS will admit dark soliton which is a defect on plane wave background. If the soliton solutions admit $\psi (x,t)= \phi (x) e^{-i \mu t}$, the NLS will be transformed to be an eigen-problem of LS with a potential term,
\begin{equation}
-\frac{1}{2} \frac{\partial^2
\phi(x)}{\partial x^2}+U(x) \phi(x) =\mu \phi(x).
\end{equation}
The potential term $U(x)=-\gamma |\phi(x)|^2 $. Especially, if there are some spatial-independent terms in the $-\gamma |\phi(x)|^2 $, the constant terms can be transferred to other side of the equation and be absorbed to the eigen-energy value.
This can be used to discuss the explicit relations between soliton solution of NLS and eigenstate of LS with a potential well.   \emph{It should be pointed out that the simple idea has been used to derive some striking solitons in NLS described systems from the solutions of LS equation\cite{Nogami,NA,Mateo,NA2}. But the relations between them have not been discussed systemically, and eigenvalues of solitons seemed to be different for attractive interactions and repulsive interactions in previous studies. In this paper, we would like to demonstrate the relations more clearly, and show more abundant solitons can be obtained from the relations.}  Many different vector solitons reported before can be also obtained from the relations directly, such as bright-bright, bright-dark, dark-bright, and dark-dark solitons \cite{Mat,Dok}. \emph{On the other hand, we will show that the soliton solution of NLS can be also used to obtain eigen-solution of LS with complicated potential wells, which is hard to be  solved directly.} Firstly, we discuss how to generate soliton solution of NLS from eigen-solutions of LS.

We will discuss them in two cases according to the attractive interaction and repulsive interaction. For $N$-component coupled  NLS with attractive interactions, we can derive many different soliton solutions from the eigen-states in some certain quantum wells. The discussions are made began with  the well-known bright soliton of the simplest scalar NLS equation with attractive interaction. For $N$-component coupled  NLS with repulsive interactions, we can derive many different soliton solutions from the identical eigen-states with the attractive cases. We discuss them started with the well-known dark soliton of the simplest scalar NLS equation with repulsive interaction. Especially, the results will show that \emph{vector solitons for coupled NLS with attractive interactions and repulsive interactions can be generated from identical quantum wells, and they admit identical eigenvalues of the quantum wells}. It should be noted that soliton with velocities can be related with static soliton through Galilean transformation, since the NLS equations admit Galilean symmetry. Therefore, we discuss  the subject mainly based on static solitons.

It should be emphasized that many of the following soliton solutions have been obtained by some other methods (Darboux transformation, Hirota bilinear forms, and other methods), and even many similar solutions have been given in \cite{Nogami,NA,Mateo,NA2}. But we would like to re-derive them and describe how to generate these soliton solution in details, to show the unified properties of solitons clearly. The parameters of soliton solutions are also written in a different way from the ones in  \cite{NA,Mateo,NA2}, to show the physical meaning of them more clearly and demonstrate the unified characters of solitons more conveniently.

\begin{figure*}[htb]
\centering
\label{fig:1}
{\includegraphics[height=105mm,width=170mm]{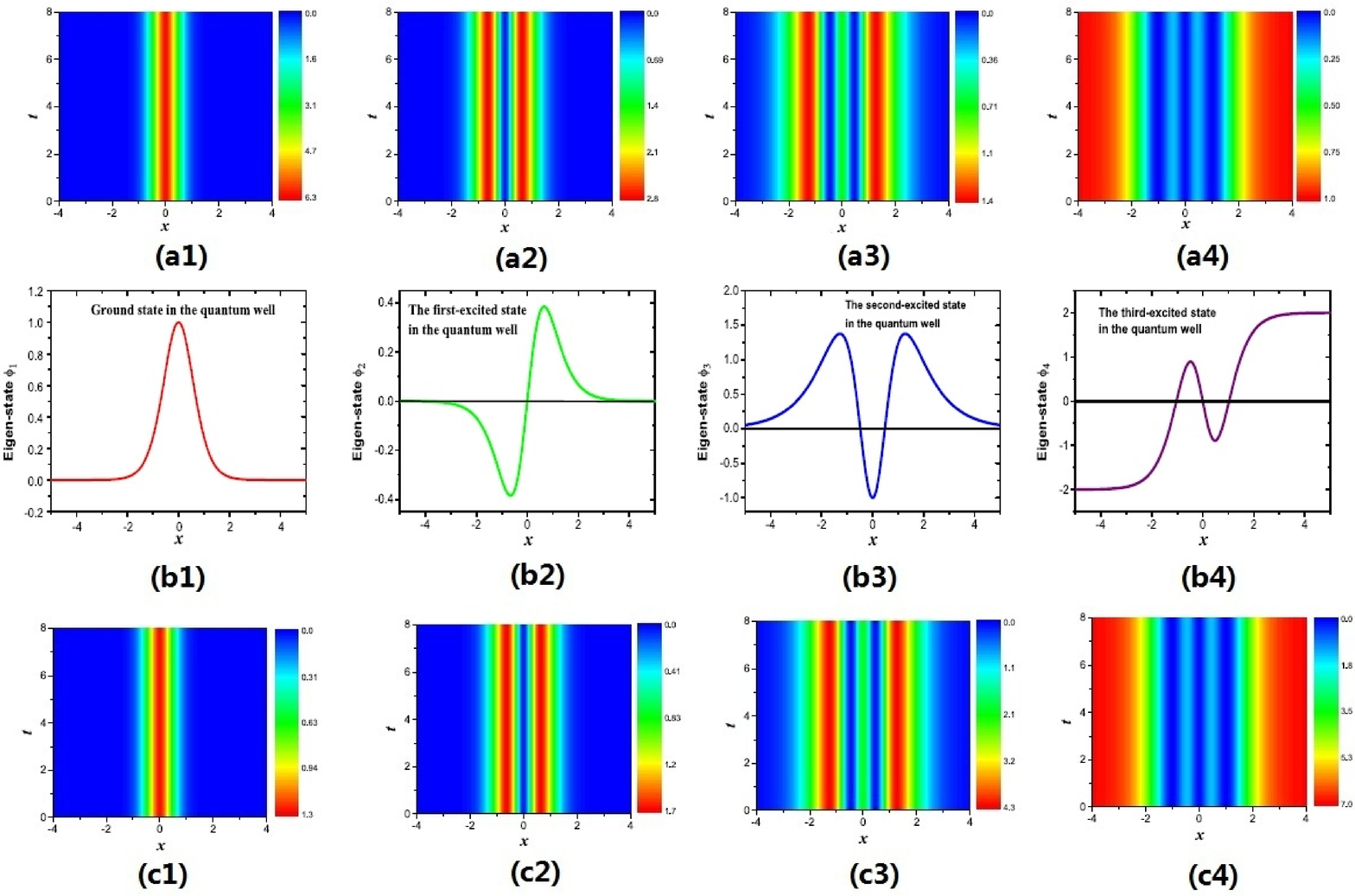}}
\caption{(color online) The correspondence between eigen-states in the quantum well and non-degenerated solitons in $4$-component coupled NLS with attractive interactions or repulsive interactions. (a1-a4) show the evolution of solitons in the four components with \textbf{attractive interactions} respectively. (b1-b4) show the \textbf{first four eigen-states in the quantum well} $-6f \sech^2[\sqrt{f} x]$. (c1-c4) show the evolution of solitons in the four components with \textbf{repulsive interactions} respectively. It is shown that the density profiles of soliton correspond to the eigen-states precisely. The solitons in attractive case have different amplitudes from the ones in repulsive cases, but they admit identical eigenvalues in the quantum well. The parameters in soliton solutions are $a=1$, and $f=1$. }
\end{figure*}

\section{Non-degenerated Soliton  of NLS with attractive interactions generated from eigen-states in quantum wells}

For $\gamma=1$ attractive case, the fundamental soliton solution of the NLS is
\begin{equation}
\psi(x,t)=\sqrt{f} \ \sech[\sqrt{f} x] \ e^{i  f/2 t},
\end{equation}
where $f>0$ denote the peak of soliton. This static soliton has been given for a long time \cite{Mat,Dok}. From the above idea, we can see that the soliton solution corresponds to the eigen-state of LS as follows,
\begin{equation}
-\frac{1}{2} \frac{\partial^2
\phi(x)}{\partial x^2}- f\   \sech^2 [\sqrt{f} x] \phi(x) =-f/2 \phi(x).
\end{equation}
$\phi(x)=\sqrt{f} \  \sech[\sqrt{f} x]$ is the corresponding eigen-state of the quantum well $- f\   \sech^2[\sqrt{f} x]$ . From the general properties of bound state in one dimensional potential in quantum theory, we can know that the eigen-state is the ground state in the quantum well, since there is no node for $\phi(x)$. The eigenvalue of energy for this eigen-state is $-f/2$. From the eigenvalues in the quantum well  $- f\   \sech^2[\sqrt{f} x]$  in \cite{Landau,Rosen}, we can know that there is the other eigenvalue which is $0$. We show them in the Table II.

 \begin{table}[!hbp]
 \centering
\begin{tabular}{|c|c|c|}
\hline
\ \ \ \  Quantum well\ \ \ \ &\ \ \ \ Eigenvalues \ \ \ \ & \ \ \ \ Eigen-states\ \  \ \ \\
\hline
\multirow{2}{*}{$-f \ \sech^2[\sqrt{f}\  x]$} & $0$ & $\tanh[\sqrt{f} x]$ \\
  & $-f/2$ & $\sech[\sqrt{f} x]$ \\
\hline
\end{tabular}
\caption{The eigenvalues and eigen-states in the quantum well $-f \ \sech^2[\sqrt{f} x]$. This can be used to construct the well-known bright-dark soliton and dark-bright soliton for two-component coupled nonlinear Sch\"{o}dinger equation. }
\end{table}

We have studied on vector soliton in multi-component coupled NLS equations. The component number can be very large and marked by $N$.  Then the $N$-component coupled NLS equation can be written as \begin{equation}\label{CNLSE}
   {\rm i}\mathbf{\psi}_t+\frac{1}{2}\mathbf{\psi}_{xx}+\gamma \mathbf{\psi}^{\dag} \mathbf{\psi}\mathbf{\psi}=0,
\end{equation}
where $\mathbf{\psi}=\left(\psi_1,\psi_2,\cdots,\psi_N\right)^T$, ``T" and ``$\dag$" represent the transpose and the Hermite conjugation of a matrix respectively.  If all components occupy the identical fundamental eigen-state $\sech[\sqrt{f} x]$, namely, $\psi_j=c_j \sqrt{f}\ \sech[\sqrt{f} x] e^{i  f/2 t}$, then the N-component coupled NLS will admit vector bright soliton. For example, the well-known bright-bright soliton can be obtained directly.  But there is an additional constrain condition on the coefficients of wave functions in components, namely, $\sum_{j=1}^N|c_j|^2=1$. When more than one component admit the same eigen-state, the corresponding soliton solutions are called by degenerated soliton. Many previous reported vector soliton, such as bright-bright \cite{Lakshman}, dark-dark soliton\cite{Lingdnls,Kivshar}, bright-bright-dark \cite{BBD}, and dark-bright-bright soliton \cite{Feng}, etc. They are all degenerated solitons \cite{DSWang}.  Since degenerated cases are just chosen from the non-degenerated soliton, we firstly focus on the non-degenerated solitons for which each component admits different eigen-state of quantum well. We will discuss degenerated soliton systemically in Section V.

From the eigen-state $tanh[\sqrt{f} x]$, we can know that there is dark soliton state, and it admits a constant background, and be a free state. Therefore, we introduce a parameter $a$ to describe the plane wave background amplitude, and assuming $|\psi_1|^2+|\psi_2|^2=a^2+f \ sech^2[\sqrt{f} x]$.
Letting $\psi_j(x,t)=\phi_j(x) e^{-i \mu_j t}$ and transferring the constant terms to eigenvalue terms, the related eigen-equations can be simplified as
\begin{eqnarray}
&&-\frac{1}{2} \frac{\partial^2
\phi_1(x)}{\partial x^2}-f\   \sech^2 [\sqrt{f} x] \phi_1(x) \nonumber\\
&& =(\mu_1 +a^2) \phi_1(x),\nonumber\\
&&-\frac{1}{2} \frac{\partial^2
\phi_2(x)}{\partial x^2}- f\   \sech^2 [\sqrt{f} x] \phi_2(x)\nonumber\\
&& =(\mu_2+a^2) \phi_2(x).
\end{eqnarray}

The eigenvalues and eigen-states for  the  quantum well $- f\   \sech^2[\sqrt{f} x]$ are given in Table II.  We can know that the fundamental eigen-values for fundamental state and the first excited state are $-f/2$ and $0$ respectively. These characters can be used to identify the phases of dark-bright soliton, namely, $\mu_1 +a^2=-f/2$ and $\mu_2 +a^2=0$.

The eigen-states correspond to $  \sech[\sqrt{f} x] $  and $ \tanh[\sqrt{f} x] $ respectively  (Table II). Therefore the eigen-state of vector  soliton   should have the same spatial distribution forms. But the coefficients of soliton can not be arbitrary, since the superposition of their density should be  $|\psi_1|^2+|\psi_2|^2=a^2+ f\   sech^2 [\sqrt{f} x] $ which is distinctive from linear cases for quantum well.

Therefore  we must introduce some new coefficient for them, namely, $\phi_1(x)= a_2 \ \sech[\sqrt{f} x] $ and  $\phi_2(x)= b_2\  \tanh[\sqrt{f} x] $.  We can identify the values of $a_2$ and $b_2$ directly with the constrain condition $|\psi_1|^2+|\psi_2|^2=a^2+ f\   \sech^2 [\sqrt{f} x] $.  Then the well-known static dark-bright soliton of the two-component NLS equation obtained as follows,
\begin{eqnarray}
\psi_1(x,t)&=&\sqrt{f+a^2} \  \sech[\sqrt{f} x] \ e^{i ( f/2+a^2) t},\nonumber\\
\psi_2(x,t)&=&a\ \tanh[\sqrt{f} x] \ e^{i a^2 t}.
\end{eqnarray}
where $a$ denotes the amplitude of plane wave background for dark soliton component.
It is seen that the bright soliton still corresponds to the eigenvalue $-f/2$ in the quantum well, and the dark soliton corresponds to eigenvalue zero in the quantum well. Especially, we can see that $\phi_2$ admit one node in spatial distribution. From the general properties of bound state in 1 dimensional potential \cite{Landau}, we know that the eigen-state of dark soliton is the first-excited state in the quantum well, and it is a free state. Therefore, we can know there is no other bound states among the ground state and free state for the quantum well $- f\   \sech^2[\sqrt{f} x]$. Even for bright-bright-dark soliton presented before \cite{BBD,Feng}, the eigen-value of the two bright solitons are the same and there is no other bound states. This comes from that all density superposition form is still a $constant - f\   \sech^2[\sqrt{f} x]$ form, the quantum well is identical with the one for two-component case. Therefore bright solitons in the two components of three-component case admit identical eigen-state of the quantum well.

\begin{table}[!hbp]
\centering
\begin{tabular}{|c|c|c|}
\hline
\ \ Quantum well\ \  &\ \  Eigenvalues \ \ & Eigen-states \\
\hline
\multirow{3}{*}{$-3 f \sech^2[\sqrt{f} x]$} & $0$ & $1-3 \tanh ^2[\sqrt{f} x]$ \\
  & $-f/2$ & $\ \ \tanh [\sqrt{f} x]\  \sech[\sqrt{f} x]$\ \ \\
   &$ -2f$ & $\sech^2[\sqrt{f} x]$ \\
\hline
\end{tabular}
\caption{The eigenvalues and eigen-states in the quantum well $-3 f \sech^2[\sqrt{f} x]$. This can be used to construct non-degenerated vector solitons for three-component coupled nonlinear Sch\"{o}dinger equation with attractive or repulsive interactions. }
\end{table}

From the knowledge that more bound states can emerge for much deeper potential well \cite{Landau}, we can obtain more bound states though making the quantum well deeper. Therefore we introduce a multiply factor $h$ to the quantum well, namely $- h\  f\   \sech^2[\sqrt{f} x]$. Based on the results in \cite{NA,Rosen} and the textbook \cite{Landau}, we can know that the parameter $h$ can not be arbitrary, and $h=N(N-1)/2$ (where $N$ is a positive integer larger than one), it is possible to present exact eigen-state analytically. Since we intend to establish the relations between soliton solution of NLS and eigen-problems of LS, therefore we focus on these integrable cases. Then, we investigate the $N=3$ case.
Letting $\psi_j(x,t)=\phi_j(x) e^{-i \mu_j t}$ and $\sum_{j=1}^3|\phi_j(x)|^2=a^2 +3 f \  \sech^2[\sqrt{f} x] $,  the related eigen-equations can be simplified as
\begin{eqnarray}
&&-\frac{1}{2} \frac{\partial^2
\phi_j(x)}{\partial x^2}-3f\   \sech^2 [\sqrt{f} x] \phi_j(x) \nonumber\\
&&=(\mu_j+a^2)\  \phi_j(x), (j=1,2,3).
\end{eqnarray}
The eigenvalues and eigen-states for  the  quantum well $- 3 f\   \sech^2[\sqrt{f} x]$ are given in Table III. We can know that the eigenvalues for ground state, the first-excited state, and the second-excited state are $-2f$,  $-f/2$ and $0$ respectively. These characters can be used to identify the phases of dark-bright-bright soliton, namely, $\mu_1 +a^2=-2f$, $\mu_2 +a^2=-f/2$ and $\mu_3 +a^2=0$.

The eigen-states correspond to $ \sech^2[\sqrt{f} x] $ ,  $ \sech[\sqrt{f} x]\ \tanh[\sqrt{f} x] $, and $(1-3 \tanh ^2[\sqrt{f} x]  $ respectively  (see the above bright-bright-dark soliton). Therefore the eigen-state of dark-bright soliton here should have the same spatial distribution forms. Similarly,  we must introduce some new coefficient for them, namely, $\phi_1(x)= a_3 \  \sech^2[\sqrt{f} x] $, $\phi_2(x)= b_3 \  \sech[\sqrt{f} x] \ \tanh[\sqrt{f} x] $, and  $\phi_3(x)=c_3(1-3 \tanh ^2[\sqrt{f} x]  $ .  We can identify the values of $a_3$ , $b_3$ and $c_3$ with the constrain condition $|\psi_1|^2+|\psi_2|^2+|\phi_3(x)|^2=a^2+3 f\   \sech^2 [\sqrt{f} x] $.  Then the static vector soliton of the three-component NLS equations obtained as follows,
\begin{eqnarray}
\psi_1(x,t)&=&\sqrt{3} \sqrt{a^2/4+f}\  \sech^2[\sqrt{f} x]\ e^{ i t (a^2+2f)},\nonumber\\
\psi_2(x,t)&=&\sqrt{3} \sqrt{ a^2+f} \tanh [\sqrt{f} x]\   \sech[\sqrt{f} x]\nonumber\\
 && e^{ i t  ( a^2+f/2 )},\nonumber\\
\psi_3(x,t)&=&\frac{a}{2}\ (1-3 \tanh ^2[\sqrt{f} x] )\  e^{ i a^2 t}.
\end{eqnarray}
We can see that there a bright soliton with one peak in component $\psi_1$, a double-peak bright soliton in component $\psi_2$, and a double-valley dark soliton in component $\psi_3$, through plotting their density evolution. The corresponding eigen-state of double-peak bright soliton admit one node in spatial distribution, and it is the first-exited state and it is also a bound state. Its eigen-value is $-f/2$. The corresponding eigen-state of double-valley dark soliton admit two nodes in spatial distribution, therefore it is the second-excited state, and it is a free state.

\begin{table}[!hbp]
\centering
\begin{tabular}{|c|c|c|}
\hline
Quantum well & Eigenvalues & Eigen-states \\
\hline
\multirow{4}{*}{$-6 f \sech^2[\sqrt{f} x]$} & $0$ & $\tanh[\sqrt{f} x]  (5 \tanh ^2[\sqrt{f} x]-3 )$ \\
  & $-f/2$ & $ (5 \tanh ^2[\sqrt{f} x]-1 )  \sech[\sqrt{f} x]$ \\
   & $-2f$  & $\tanh[\sqrt{f} x] \ \sech^2[\sqrt{f} x]$ \\
   &$ -9f/2$ & $\sech^3 [\sqrt{f} x]$ \\
\hline
\end{tabular}
\caption{The eigenvalues and eigen-states in the quantum well $-6f \sech^2[\sqrt{f} x]$. This can be used to construct non-degenerated vector solitons for four-component coupled nonlinear Sch\"{o}dinger equation with attractive or repulsive interactions. }
\end{table}

Then, we investigate the $N=4$ case. Letting $\psi_j(x,t)=\phi_j(x) e^{-i \mu_j t}$ and $\sum_{j=1}^4|\phi_j(x)|^2=a^2 +6 f \  \sech^2[\sqrt{f} x] $,  the vector soliton solution of 4-component coupled NLS  can be derived in a similar way. The eigenvalues and eigen-states are given in Table IV. It is seen that the eigenvalue of fundamental state is $-9f/2$ for the quantum well $-6 f \  \sech^2[\sqrt{f} x] $. The vector soliton solution is derived as follows,
\begin{eqnarray}
\psi_1(x,t)&=&\frac{1}{2} \sqrt{\frac{5}{2}(a^2+9 f)}  \sech^3 [\sqrt{f} x]\  e^{\frac{1}{2} i t (2 a^2+9 f )},\nonumber\\
\psi_2(x,t)&=&\frac{1}{2} \sqrt{15(a^2+4 f)} \tanh[\sqrt{f} x] \ \sech^2[\sqrt{f} x]\nonumber\\
 &&  e^{i t  (a^2+2 f )},\nonumber\\
\psi_3(x,t)&=&\frac{1}{2} \sqrt{\frac{3}{2}(a^2+f)}   (5 \tanh ^2[\sqrt{f} x]-1 )  \sech[\sqrt{f} x]\nonumber\\
 && e^{\frac{1}{2} i t (2 a^2+f )} ,\nonumber\\
\psi_4(x,t)&=&\frac{1}{2} a \tanh[\sqrt{f} x]  (5 \tanh ^2[\sqrt{f} x]-3 )\  e^{i a^2 t}.
\end{eqnarray}
We can see that there a bright soliton with one peak in component $\psi_1$, a double-peak bright soliton in component $\psi_2$, triple-peak bright soliton in component $\psi_3$, and a triple-valley dark soliton in component $\psi_4$, through plotting their density evolution.  The results are shown in Fig. 1 (a). The corresponding eigen-state of double-peak bright soliton admit one node in spatial distribution, and it is the first-exited state and its eigenvalue is $-2f$. The corresponding eigen-state of triple-peak bright soliton admit two nodes in spatial distribution, therefore it is the second-excited state and its eigenvalue is $-f/2$. The triple-valley dark soliton admit three nodes and it is the third-excited state and its eigenvalue still be zero. The nodes character of them are shown in Fig. 1 (b).

 It should be pointed that the above results are identical with the results in \cite{NA}. Here we demonstrate more clearly that the solitons of NLS can be related with eigen-state in $ -h f\ \sech ^2[\sqrt{f} x]$ type quantum wells.  But we here further show that the eigenvalues and eigen-state can be also used to construct soliton solutions of NLS with repulsive interactions. As far as we know, this has not been discussed systemically before.

\section{Non-degenerated Soliton  of NLS with repulsive interactions generated from eign-states in quantum wells}
For $\gamma=-1$ repulsive case, the fundamental soliton solution of the NLS is
\begin{equation}
\psi(x,t)=\sqrt{f} \ \tanh[\sqrt{f} x] \ e^{-i  f t},
\end{equation}
where $f>0$ denote amplitude of plane wave background for dark soliton. This static dark soliton has been given for a long time \cite{Zk}, and it has been observed in many experiments \cite{darksoliton1,darksoliton2}. Substituting it to NLS and we can obtain similar eigen-problem in a quantum potential,
\begin{equation}
-\frac{1}{2} \frac{\partial^2
\phi(x)}{\partial x^2}+ f\   \tanh^2 [\sqrt{f} x] \phi(x) =f \phi(x).
\end{equation}
It looks like that the eigenvalue of dark soliton is $f>0$ in this form. The potential is $f \   \tanh^2 [\sqrt{f} x]$. But the potential form can be rewritten as $f-f\   \sech^2 [\sqrt{f} x]$ with the aid of $\tanh^2(x)=1- \sech^2(x)$. In this way we can also make a correspondence between dark soliton in repulsive case with the eigen-problem in a identical quantum well  $-f\   \sech^2 [\sqrt{f} x]$ with the attractive case. Namely,
\begin{equation}
-\frac{1}{2} \frac{\partial^2
\phi(x)}{\partial x^2}- f\  \sech^2 [\sqrt{f} x] \phi(x) =0 \phi(x).
\end{equation}
 Therefore, the dark soliton in repulsive case also corresponds to zero eigen-value of energy in the quantum well  $-f\   \sech^2 [\sqrt{f} x]$ . Then, we can find the fundamental state with the help of dark-bright soliton solutions in two-component coupled NLS with repulsive interactions. This is different from the case with attractive interactions, since we obtain the first excited state and try to find the ground state for the repulsive case.

 Comparing with the $|\psi_1|^2+|\psi_2|^2=a^2+f \ \sech^2[\sqrt{f} x]$ form  in two-component coupled NLS with attractive case, we introduce a form to describe potential wells, $a^2+ f\   \tanh^2 [\sqrt{f} x] = |\psi_1|^2+|\psi_2|^2$  to find the ground state in the repulsive cases. With the help of $\tanh^2[x]=1- \sech^2[x]$, we can rewrite the potential form as  $a^2+ f\   \tanh^2 [\sqrt{f} x] =a^2+f\ -f\ \sech^2[\sqrt{f} x] $. Similar to the case in attractive case, letting $\psi_j(x,t)=\phi_j(x) e^{-i \mu_j t}$ and transferring the constant terms to eigenvalue terms, the related eigen-equations can be simplified as
\begin{eqnarray}
&&-\frac{1}{2} \frac{\partial^2
\phi_1(x)}{\partial x^2}-f\   \sech^2 [\sqrt{f} x] \phi_1(x) \nonumber\\
&& =(\mu_1 -a^2-f) \phi_1(x),\nonumber\\
&&-\frac{1}{2} \frac{\partial^2
\phi_2(x)}{\partial x^2}- f\  \sech^2 [\sqrt{f} x] \phi_2(x)\nonumber\\
&& =(\mu_2-a^2-f) \phi_2(x).
\end{eqnarray}

 For the results for the  quantum well $- f\   \sech^2[\sqrt{f} x]$, we can know that the  eigenvalues for ground state and the first-excited state are $-f/2$ and $0$ respectively. These characters can be used to identify the phases of dark-bright soliton, namely, $\mu_1 -a^2-f=-f/2$ and $\mu_2 -a^2-f=0$.
The eigen-states correspond to $  \sech[\sqrt{f} x] $  and $ \tanh[\sqrt{f} x] $ respectively  (see the above bright-dark soliton). Therefore the eigen-state of dark-bright soliton here should have the same spatial distribution forms. But the coefficients of the form can not be the same, since the superposition of their density should be  $|\psi_1|^2+|\psi_2|^2=a^2+ f\   \tanh^2 [\sqrt{f} x] $ which is distinctive from the attractive cases.

Therefore  we must introduce some new coefficient for them, namely, $\phi_1(x)= a_2 \  \sech[\sqrt{f} x] $ and  $\phi_2(x)= b_2\  \tanh[\sqrt{f} x] $.  We can identify the values of $a_2$ and $b_2$ directly with the constrain condition $|\psi_1|^2+|\psi_2|^2=a^2+ f\  \tanh^2 [\sqrt{f} x] $.  Then the well-known static dark-bright soliton of the two-component NLS equation obtained as follows,
\begin{eqnarray}
\psi_1(x,t)&=& a \  \sech[\sqrt{f} x] \ e^{i ( f/2-a^2-f) t},\nonumber\\
\psi_2(x,t)&=&\sqrt{f+a^2} \ \tanh[\sqrt{f} x] \ e^{-i (a^2+f) t},
\end{eqnarray}
where $\sqrt{f+a^2}$ denotes the amplitude of plane wave background for dark soliton component. It is shown that the dark-bright soliton form for repulsive interactions is distinctive from the bright-dark soliton for attractive cases. But they are related with the eigenvalue in the same quantum well $-f\ \sech^2[\sqrt{f} x] $. This provides a possible way to construct static vector soliton with repulsive case from the vector soliton with attractive interactions.

\begin{table}[!hbp]
\centering
\begin{tabular}{|c|c|c|}
\hline
Quantum well & Eigenvalues & Soliton in NLS \\
\hline
\multirow{4}{*}{$-6 f \sech^2[\sqrt{f} x]$} & $0$ & triple-valley dark soliton \\
  & $-f/2$ & triple-hump bright soliton \\
   & $-2f $ & double-hump bright soliton \\
   & $-9f/2$ & single-hump bright soliton \\
\hline
\end{tabular}
\caption{The correspondence between solions in $4$-component coupled NLS with repulsive interactions and  the eigenvalues in quantum well $-6f \sech^2[\sqrt{f} x]$. This can be used to construct degenerated vector solitons for $N$-component($N>4$) coupled nonlinear Sch\"{o}dinger equation with repulsive interactions. }
\end{table}
Similar to the attractive case for more eigen bound states, we further deep the potential well $a^2+3  f\   \tanh^2 [\sqrt{f} x] = |\psi_1|^2+|\psi_2|^2+|\psi_3|^2$. Letting $\psi_j(x,t)=\phi_j(x) e^{-i \mu_j t}$ and transferring the constant terms to eigenvalue terms, the related eigen-equations can be simplified as
\begin{eqnarray}
&&-\frac{1}{2} \frac{\partial^2
\phi_j(x)}{\partial x^2}-3f\   \sech^2 [\sqrt{f} x] \phi_1(x) \nonumber\\
&&=(\mu_j-a^2-3 f)\  \phi_j(x).  (j=1,2,3)
\end{eqnarray}
For the above results for the  quantum well $- 3 f\   \sech^2[\sqrt{f} x]$, we can know that the fundamental eigen-values for fundamental state, the first excited state, and the second excited state are $-2f$,  $-f/2$ and $0$ respectively. These characters can be used to identify the phases of dark-bright-bright soliton, namely, $\mu_1 -a^2-3f=-2f$, $\mu_2 -a^2-3f=-f/2$ and $\mu_3 -a^2-3f=0$.

The eigen-states correspond to $ \sech^2[\sqrt{f} x] $ ,  $ \sech[\sqrt{f} x]\ \tanh[\sqrt{f} x] $, and $(1-3 \tanh ^2[\sqrt{f} x]  $ respectively  (see the above three-component vector soliton). Therefore the eigen-state of vector soliton for repulsive interactions should have the same spatial distribution forms. Similarly,  we must introduce some new coefficient for them, namely, $\phi_1(x)= a_3 \  \sech^2[\sqrt{f} x] $, $\phi_2(x)= b_3 \  \sech[\sqrt{f} x] \ \tanh[\sqrt{f} x] $, and  $\phi_3(x)=c_3(1-3 \tanh ^2[\sqrt{f} x]  $ .  We can identify the values of $a_3$ , $b_3$ and $c_3$ with the constrain condition $|\psi_1|^2+|\psi_2|^2+|\phi_3(x)|^2=a^2+3 f\   \tanh^2 [\sqrt{f} x] $.  Then the static vector soliton of the three-component NLS equations obtained as follows,
\begin{eqnarray}
\psi_1(x,t)&=&\frac{1}{2}\sqrt{ 3 (a^2-f)} \  \sech^2[\sqrt{f} x]\ e^{-i t  (a^2+f )},\nonumber\\
\psi_2(x,t)&=&\sqrt{3  (a^2+2 f )}  \tanh [\sqrt{f} x]\   \sech[\sqrt{f} x]\nonumber\\
 && e^{-i t  (a^2+\frac{5 f}{2} )},\nonumber\\
\psi_3(x,t)&=&\frac{1}{2}\sqrt{  (a^2+3 f)}\ (1-3 \tanh ^2[\sqrt{f} x] )\nonumber\\
 &&  e^{-i t  (a^2+3 f )}.
\end{eqnarray}
We emphasize that this solution is quite distinctive from the ones reported before \cite{BBD,Feng}.  The solutions are similar to the ones in \cite{darkwt}.

In a similar way, we can obtain the vector soliton in four-component coupled NLS with repulsive case as follows from the quantum well further deep the potential well $a^2+6  f\   \tanh^2 [\sqrt{f} x] = |\psi_1|^2+|\psi_2|^2+|\psi_3|^2+|\psi_4|^2$,
\begin{eqnarray}
\psi_1(x,t)&=&\frac{1}{2} \sqrt{\frac{5}{2}  (a^2-3 f )}   \sech^3 [\sqrt{f} x]\  e^{- i(a^2+3f/2)t},\nonumber\\
\psi_2(x,t)&=& \frac{1}{2} \sqrt{15  (a^2+2 f )} \tanh[\sqrt{f} x] \ \sech^2[\sqrt{f} x]\nonumber\\
 &&  e^{-i(a^2+4f)t },\nonumber\\
\psi_3(x,t)&=&\frac{1}{2} \sqrt{\frac{3}{2} (a^2+5 f )}    (5 \tanh ^2[\sqrt{f} x]-1 )  \sech[\sqrt{f} x]\nonumber\\
 && e^{-i(a^2+11f/2)t } ,\nonumber\\
\psi_4(x,t)&=&\frac{1}{2} \sqrt{a^2+6 f}  \tanh[\sqrt{f} x]  (5 \tanh ^2[\sqrt{f} x]-3 )\nonumber\\
 &&  e^{-i (a^2+6f) t}.
\end{eqnarray}
The density evolution is shown in Fig. 1(c). The corresponding eigen-state is shown in Fig. 1(b). The correspondence between soliton excitation and eigenvalues of quantum well are summarized in Table V .

Furthermore, we  establish the correspondence between non-degenerated solitons and eigen-states in the quantum well $-N(N-1)/2 f \ \sech[\sqrt{f} x]$ to generalized $N$-component coupled NLS equation with attractive or repulsive interactions. The results is summarized in Table I. This is the main results in this paper.

It should be pointed out that the many soliton solutions re-derived from the eigen-states had been presented by Darboux-transformation, Hirota bilinear method, which usually admit $\sech^1(x)$ form even for multi-component coupled cases. Also we emphasize that many multi-soliton complexes had been given in \cite{NA,NA2}, which were also derived from the solutions of a LS with adding some constrain conditions. But the correspondence between eigen-states of LS and NLS with attractive or repulsive interactions have not been addressed systemically. Especially, we show that solitons of the arbitrary $N$-component NLS with repulsive interactions have identical eigenvalues with the ones of $N$-component NLS with attractive interactions. Furthermore, there is a special characters for dark solitons of $N$-component NLS with both attractive and repulsive interactions, namely,\emph{ they always correspond to the zero energy  eigenvalue in the quantum well, with a proper reference for which free plane wave admits zero kinetic energy}. These underlying characters have not been uncovered clearly in previous studies, and they enable us to obtain soliton solution of $N$-component coupled NLS with repulsive interactions from the ones in NLS with  attractive interactions. It is well known that Darboux transformation forms are different for the NLS with repulsive and attractive interactions \cite{Lingdnls}. Some authors even try to derive dark solitons by Hirota method \cite{Kivshar,Feng}, or the KP-hierarchy reduction method \cite{DSWang}. Therefore, these underlying characters are non-trivial and enable us to derive vector solitons
more conveniently. Based on the non-degenerated soliton presented above, we can obtain very abundant vector solitons directly, such as degenerated solitons, and beating solitons.

\section{Degenerated vector solitons generated from non-degenerated solitons}
From the above non-degenerated soliton solution, we can know that soliton solution of $M$-component coupled NLS equation can be used to construct much abundant vector solitons in $N$-component coupled cases with $N>M$. Obviously, larger $M$ component number will generate more abundant vector soliton types. As examples, we show explicitly how many vector solitons for $4$-component coupled cases can be generated from the $M$ component cases with $M\in [1,3]$.

\emph{$4$-component degenerated soliton generated from the scalar NLS. } There are mainly two cases for scalar NLS. It admits bright soliton with attractive interactions, and dark soliton with repulsive interactions. For linear quantum LS, it is obviously the coefficient of eigen-state can be arbitrary nonzero. But there is a constrain condition for degenerated soliton in multi-component case. The simplest degenerated soliton is a vector bright soliton for 4-component NLS with attractive interactions, which can be written as $\psi_j=c_j \sqrt{f} \ \sech[\sqrt{f} x] \ e^{i  f/2 t}$, with the constrain condition $\sum_{j=1}^4|c_j|^2=1 $. Similarly, we can directly write the degenerated vector dark soliton for 4-component NLS with repulsive interactions,  $\psi_j=c_j \sqrt{f} \ \tanh[\sqrt{f} x] \ e^{-i  f t}$, with the constrain condition $\sum_{j=1}^4|c_j|^2=1 $. Obviously, the vector bright soliton can be extended to N-component coupled case, which agree well with N-bright-bright soliton given in \cite{Lakshman}, and
N-dark-dark solitons in the integrable coupled NLS equations derived by
the KP-hierarchy reduction method \cite{DSWang}.

\emph{$4$-component degenerated soliton generated from the non-degenerated soliton in two-component NLS.}
The two-component NLS admit bright-dark soliton for attractive interactions, and dark-bright soliton for repulsive interactions. Therefore, for 4-component coupled NLS with attractive interactions, there are many different degenerated solitons since there are many different degenerated numbers an degenerated states. If three components  degenerate in fundamental eigen-state, and other one component do not degenerate, then the vector soliton will become
$\psi_j=c_j\sqrt{f+a^2} \  \sech[\sqrt{f} x] \ e^{i ( f/2+a^2) t}, (j=1,2,3)$, and
$\psi_4= a\ \tanh[\sqrt{f} x] \ e^{i a^2 t}$. Since the superposition of them should be $a^2+ f\   \sech^2 [\sqrt{f} x]$, the simple constrain conditions can be chosen as $|c_1|^2+|c_2|^2+|c_3|^2=1$. If two components  degenerate in fundamental eigen-state, and other two components degenerate in the first excited state, then the vector soliton will become
$\psi_j=c_j\sqrt{f+a^2} \  \sech[\sqrt{f} x] \ e^{i ( f/2+a^2) t}, (j=1,2)$, and
$\psi_j=c_j a\ \tanh[\sqrt{f} x] \ e^{i a^2 t}, (j=3,4)$, with the simple constrain conditions  $|c_1|^2+|c_2|^2=1$, and $|c_3|^2+|c_4|^2=1$.   If three components  degenerate in the first excited eigen-state, and other one component do not degenerate, then the vector soliton will become
$\psi_1=\sqrt{f+a^2} \  \sech[\sqrt{f} x] \ e^{i ( f/2+a^2) t},$, and
$\psi_j=c_j a\ \tanh[\sqrt{f} x] \ e^{i a^2 t}$, (j=2,3,4), with the simple constrain conditions  $|c_2|^2+|c_3|^2+|c_4|^2=1$. These simple constrain conditions satisfy that the coefficients of degenerated states are summarized to be unity. Similarly, we can write out the relation expressions for vector solitons for which three components degenerate in dark soliton, and one component occupy bright soliton. The cases for 4-component coupled NLS with repulsive interactions can be written out from the dark-bright soliton presented above in a similar way. We do not show them in details anymore. We emphasize that all these soliton solutions generated from scalar NLS or two-component NLS have been also derived by Darboux transformation or other methods \cite{Lakshman,Feng,Lingdnls,Kivshar,DSWang}.

\emph{$4$-component degenerated soliton generated from the non-degenerated solitons in  three-component NLS.} There are mainly three different cases for degenerated soliton in 4-component NLS which are generated from the non-degenerated solitons in three-component NLS, since there are at most two degenerated components in this case. Firstly, if there are two components degenerated in fundamental state, then the vector solitons for repulsive interaction case can be written as
$  \psi_j=c_j \frac{1}{2}\sqrt{ 3 (a^2-f)} \  \sech^2[\sqrt{f} x]\ e^{-i t  (a^2+f )}, (j=1,2)$,
$\psi_3 =\sqrt{3  (a^2+2 f )}  \tanh [\sqrt{f} x]\   \sech[\sqrt{f} x]\ e^{-i t  (a^2+\frac{5 f}{2} )}$,
and $\psi_4 =\frac{1}{2}\sqrt{  (a^2+3 f)}\ (1-3 \tanh ^2[\sqrt{f} x] )\  e^{-i t  (a^2+3 f )}$. The constrain condition is $|c_1|^2+|c_2|^2=1$, since the superposition of them must be $a^2+3 f\   \sech^2 [\sqrt{f} x]$ . There are three bright solitons in three components separately and one dark soliton in the fourth component, which is similar to the case generated from the dark-bright soliton in two-component case. But the soliton profiles are different from the ones obtained above, namely, there are double-hump soliton and double-valley dark soliton in this case. Secondly,  if there are two components degenerated in the first excited state, then the vector solitons for repulsive interaction case can be written as
$  \psi_1= \frac{1}{2}\sqrt{ 3 (a^2-f)} \  \sech^2[\sqrt{f} x]\ e^{-i t  (a^2+f )}, $
$\psi_j =c_j \sqrt{3  (a^2+2 f )}  \tanh [\sqrt{f} x]\   \sech[\sqrt{f} x]\ e^{-i t  (a^2+\frac{5 f}{2} )}, (j=2,3)$,
and $\psi_4 =\frac{1}{2}\sqrt{  (a^2+3 f)}\ (1-3 \tanh ^2[\sqrt{f} x] )\  e^{-i t  (a^2+3 f )}$, with the constrain condition  $|c_2|^2+|c_3|^2=1$. Thirdly,  if there are two components degenerated in the second excited state, then the vector solitons for repulsive interaction case can be written as
$  \psi_1= \frac{1}{2}\sqrt{ 3 (a^2-f)} \  \sech^2[\sqrt{f} x]\ e^{-i t  (a^2+f )}, $
$\psi_2=\sqrt{3  (a^2+2 f )}  \tanh [\sqrt{f} x]\   \sech[\sqrt{f} x]\ e^{-i t  (a^2+\frac{5 f}{2} )}$,
and $\psi_j =c_j \frac{1}{2}\sqrt{  (a^2+3 f)}\ (1-3 \tanh ^2[\sqrt{f} x] )\  e^{-i t  (a^2+3 f )}, (j=3,4)$, with the constrain condition  $|c_3|^2+|c_4|^2=1$. The third case corresponds to that there are one bright soliton with single hump in one component, one bright soliton with two humps in one component, and two dark soliton with two valleys in the other two components separately. As far as we know, these types of soliton have not been derived by Darboux transformation and other methods \cite{Lakshman,Feng,Lingdnls,Kivshar,DSWang}, and even in \cite{NA,NA2}.

It is well known that the arbitrary linear superposition of eigen-states of LS equation with the quantum well are still the solution of the LS equation. Therefore the linear superposition of the above non-degenerated and degenerated soliton solutions will also be used to generated localized waves. However, there are some constrain conditions on the superposition coefficients for NLS equation, since the effective potential in NLS depend on the wave functions. The linear superposition forms are generated from these eigen-states, but they demonstrate striking dynamics characters. The beating dark-dark soliton was obtained by the similar $SU(2)$ symmetry \cite{park,Yanbd}. Based on the above results, we can know that the beating effects are induced by the eigen-value difference between solitons \cite{Zhao5}.   In fact, this simple idea can be used to generated very abundant localized waves with beating effects. Explicitly, $N$-component coupled NLS with equal nonlinear coefficients admit $SU(N)$ symmetry, and related unitary matrix can be used to construct beating solitons. Especially, the forms of unitary matrix can be chosen to admit $SU(M)$ ($M\leq N$) symmetry,  which enables us to obtain very abundant different beating patterns \cite{Zhao5}.

\textbf{The above solutions are all derived from the eigen-states in quantum wells}. The corresponding relations deep our understanding on the eigenvalues of solitons and the relations between solitons in attractive and repulsive interaction cases. Then, we would like to show soliton solution of NLS could also help us to obtain eigen-state and eigenvalues in more complicated quantum wells.

\begin{figure}[htb]
\centering
\label{fig:2}
{\includegraphics[height=60mm,width=85mm]{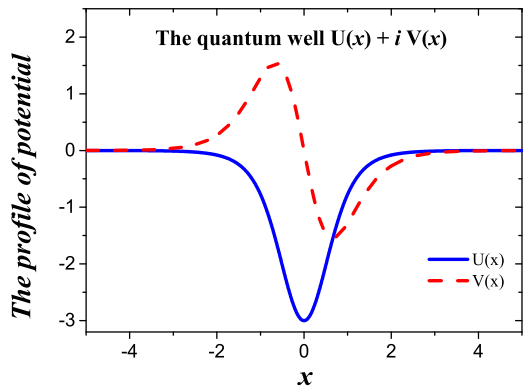}}
\caption{(color online)  The profiles of quantum well which makes the Hamiltonian belong to  the non-Hermitian Hamiltonian having Parity-Time symmetry. The quantum well admits the profile as $U(x)=-f \ \sech^2[\sqrt{f} x] (2 \beta^2 f \ \sech^2[\sqrt{f} x]+1)$ and $V(x)= - 4 \beta f^{3/2} \tanh [\sqrt{f} x] \ \sech^2[\sqrt{f} x] $, for which it is hard to solve the Hamiltonian directly. But the corresponding eigen-state and eigenvalue can be obtained from the soliton solution of a nonlinear partial equation which is solved by Darboux transformation.  The parameters  are $\beta=1$, and $f=1$. }
\end{figure}

\section{Eigen-state in a complicated quantum well generated  from soliton excitation of Nonlinear equations}
We consider the following LS equation with a more complicated well,
\begin{eqnarray}
&&-\frac{1}{2} \frac{\partial^2
\phi(x)}{\partial x^2}+[U(x)+i V(x)] \phi(x)=\mu \phi(x),
\end{eqnarray}
where $U(x)=-f \ \sech^2[\sqrt{f} x] (2 \beta^2 f \ \sech^2[\sqrt{f} x]+1)$ and $V(x)= - 4 \beta f^{3/2} \tanh [\sqrt{f} x] \ \sech^2[\sqrt{f} x] $. The profiles of them are shown in Fig. 4. Obviously, the corresponding Hamiltonian belongs to  the non-Hermitian Hamiltonian having Parity-Time symmetry as defined in \cite{PTs,JYang,Konotop}.  It is hard to solve it directly, since the quantum well is much more complicated than the ones discussed above. We note that the eigenvalue problem can be related with a NLS with some high-order effects (the well-known  Kundu-Eckhaus   equation) \cite{K-E}, namely,  $i\frac{\partial \psi(x,t)}{\partial t}+\frac{1}{2} \frac{\partial^2
\psi(x,t)}{\partial x^2}+ |\psi(x,t)|^2 \psi(x,t)  +2\beta^2 |\psi(x,t)|^4 \psi(x,t) -2i  \beta |\psi(x,t)|_x \psi(x,t)=0$. Based on the soliton solution derived by Darboux transformation method \cite{Geng}, we can represent the soliton as $\psi(x,t)= \sqrt{f}  \sech[\sqrt{f} x] \exp \left(i 2 \beta \sqrt{f} \exp (\sqrt{f} x) \sech[\sqrt{f} x]+\frac{i f t}{2}\right)$. Then we can obtain the eigenvalue of the Parity-Time symmetry potential is $\mu=-f/2$, and the corresponding eigen-state is
\begin{eqnarray}
\phi(x)&=&  \sqrt{f}  \sech[\sqrt{f} x] \nonumber\\&&\  \exp \left[i 2 \beta \sqrt{f} \exp (\sqrt{f} x) \sech[\sqrt{f} x]\right].
\end{eqnarray}
We can see clearly that the non-Hermitian Hamiltonian having Parity-Time symmetry admit real eigenvalue. The soliton state is the fundamental eigen-state in the quantum well. The excited states need more systemical analysis with coupled models. We will discuss on this subject in a separated literature. The similar discussions can be made to obtain more complicated quantum wells, even  for NLS with Parity-Time symmetry potential cases \cite{PTyan, yan2}.

\begin{figure}[htb]
\centering
\label{fig:2}
{\includegraphics[height=60mm,width=85mm]{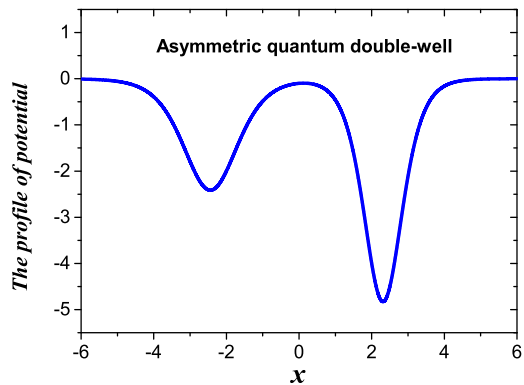}}
\caption{(color online)  The profiles of an asymmetric quantum double-well, for which it is hard to solve the Hamiltonian directly. But the corresponding eigen-state and eigenvalue can be obtained from the soliton solution of a nonlinear partial equation which is solved by Darboux transformation.  The parameters  are $f=1$, and $h=2$. }
\end{figure}

Moreover, the asymmetric soliton solution of coupled NLS reported in \cite{NA3} can be used to construct eigen-states of asymmetric quantum double-well. For an example, we show a simple case to present the ground state and the first exited state in an asymmetric quantum double-well. The corresponding quantum well admits the following form
\begin{eqnarray}
&&-\frac{1}{2} \frac{\partial^2
\phi(x)}{\partial x^2}+[V_{dw}(x) ] \phi(x)=\mu \phi(x),
\end{eqnarray}
where $V_{dw}(x) =-\frac{2  (\sqrt{f}+\sqrt{f+1} ) }   { (\sqrt{f+1}-\sqrt{f} ) H^2(x)} [f \cosh [2 \sqrt{f+1} (x-h)]+(f+1) \cosh [2 \sqrt{f} (h+x)]+1]$  describes a double-well with variable profiles. The $ H(x)=\cosh [\sqrt{f+1} (x-h)+\sqrt{f} (h+x)]+\sqrt{\frac{\sqrt{f}+\sqrt{f+1}}{\sqrt{f+1}-\sqrt{f}}} \cosh [\sqrt{f+1} (x-h)-\sqrt{f} (h+x)] $. When $f=1$ and $h=2$, the quantum potential form will admits asymmetric double-well, and it is shown in Fig. 4. It is hard to solve the LS with this potential. But the eigen-states can be given from soliton states in  \cite{NA3}, namely,
\begin{eqnarray}
\phi_1(x)&=& \frac{2 \sqrt{f+1} \sqrt{\frac{\sqrt{f}+\sqrt{f+1}}{\sqrt{f+1}-\sqrt{f}}} \cosh \left(\sqrt{f} (h+x)\right)}{H(x) }, \\
\phi_2(x)&=& \frac{2 \sqrt{f} \sqrt{\frac{\sqrt{f}+\sqrt{f+1}}{\sqrt{f+1}-\sqrt{f}}} \sinh \left(\sqrt{f+1} (x-h)\right)}{H(x)},
\end{eqnarray}
which are the ground state and the first-excited state in the quantum well. The eigenvalues of the eigen-states are $-\frac{f+1} {2}$ and $-\frac{f} {2}$ respectively.  These results show clearly that the simple relations between solitons and eigen-states in quantum well indeed can help us to solve some complicated problems.

\section{Conclusion and discussion}
In this way,  a systemic relation is established  between the eigen-state in quantum wells and the soliton solution of multi-component coupled NLS with attractive or repulsive interactions. It is shown that static soliton excitation of $N$-component coupled nonlinear Sch\"{o}dinger equation corresponds to the first $N$ eigen-states in a $-\frac{N(N-1)}{2}  f\  \sech ^2(f \ x)$ type potentials. The corresponding  energy eigenvalue of them are calculated exactly. Explicitly, $N-1$ soliton states on zero background have negative energy eigenvalues and are all bound states, the dark solitons always correspond to the eigen-states with zero energy eigenvalue in the quantum wells. \emph{The characters holds well for both attractive and repulsive nonlinear interaction cases}. The above results are summarized in the Table I.   These results enable us to understand the soliton solution of NLS with attractive and repulsive interactions in a unified way.

Abundant degenerated soliton can be derived directly from the non-degenerated soliton solutions. Degenerated soliton for $N$-component coupled NLS with attractive or repulsive interactions can be generated from the non-degenerated solitons of $M$-component coupled NLS, even with $N>M$. Moreover, the linear superposition principle in quantum mechanics can be used to construct very abundant vector soliton with beating effects in the systems with some certain constrain conditions on the superposition coefficients. The beating behavior of solitons can be also extended to the high-dimensional systems \cite{Charalampidis,wangwl}.

 \emph{Although the above soliton solutions are all static, but related solitons with moving velocities can be obtained directly by Galilean transformation for the NLS equations admit Galilean symmetry.} We emphasize that the analytical solution describing interactions between these soltions can not be derived from the eigen-states in quantum wells. But these solitons with different moving velocities can be chosen as initial states to simulate soliton interaction numerically, since when solitons are well separated, they can be seen as the linear superposition of solitons with different velocities approximately \cite{zhaofer,PAG}. For examples, the collision between solitons with nodes were   investigated in \cite{Kutuzov}, which demonstrated they were robust to collisions and even noises.  Additionally, a complicated method was proposed to study the collision of these solitons with nodes analytically \cite{NA3}, with the aid of results in \cite{Nogami}. We do not discussed this subject in this paper. As far as we know, the solitons with more than one nodes have not been derived by the well-known Darboux transformation method \cite{Mat,Dok,Lingdnls}, Hirota bilinear method \cite{Lakshman,Feng,Kivshar}, and even other methods \cite{DSWang}. Some of our results are similar to the soliton solutions with more than one nodes presented before \cite{NA,NA2}, which were also generated from the eigen-states of LS. However, we focus on the correspondence relations between solitons in nonlinear systems and eigen-states in quantum wells.  The degenerated solitons or beating solitons generated from the eigen-states have also not been addressed systemically before.

On the other hand, we also demonstrate soliton solution of nonlinear partial equation can be  used to solve the eigen-problems of quantum wells. As an example, we present eigenvalue and its eigen-state in a complicated quantum well for which the Hamiltonian belongs to  the non-Hermitian Hamiltonian having Parity-Time symmetry.  We further present the ground state and the first exited state in an asymmetric quantum double-well from asymmetric solitons. This paves the ways to solve the LS with complicated potentials with the aid of well-developed techniques for solving NLS equations.  This is very meaningful for LS with more complicated quantum wells which is hard to be solved directly, since the exact solutions of LS equation with different potentials play an important role in mathematics and physics \cite{Wen,Zhang,Agboola}.   These relations suggest that the stability of soliton states is related with the evolution of eigen-states in quantum wells directly. The stability of moving solitons can be also understood by the eigen-states in moving localized quantum wells. This provides another different way to understand the stability of solitons in NLS described systems.

  As an example, we discuss the possibilities to observe these quantum states in BEC systems \cite{XBO}.  Soliton interaction in BEC systems have been demonstrated widely in real experiments \cite{Billam,Nguyen,Marchant,Medley}. The experimental techniques can  be used to produce the initial conditions for observing these soliton states, and the bound states or qusi-bound states have been shown widely to  be robust. Very recently, dark-bright-bright soliton was experimentally demonstrated in spinor BEC systems \cite{Bersano}. It is noted that the dark soliton component admits one node and the two bright soltion components admit identical eigen-state, and they are degenerated. Therefore the observed solitons correspond to the degenerated soliton case here. In fact, the vector soliton in the three-component coupled systems can admit non-degenerated solitons which are presented above. The dark soliton in one component can admit double-valley structure, and one bright soliton component can also admit a node which makes the bright soliton admits double-hump. These are quite different from the ones observed in \cite{Bersano}. Therefore, the results  here predict much more abundant soliton profiles in coupled NLS described systems \cite{Coskun}. 
  Many different types of beating solitons were also demonstrated and more beating patterns are expected \cite{Zhao5}. The results here can be used to construct more exotic beating solitons with the help of symmetry properties. The evolution of quantum states could be observed in some classical systems, such as nonlinear fiber, water wave tank, etc., since the NLS describe many different classical systems \cite{Onorato} and  a generalized hydrodynamics could provide a remarkable quantum-classical equivalence \cite{Doyon}.

\section*{Acknowledgments}
 Zhao is grateful to C.M. Guo for her helpful discussions on plotting figures. This work is supported by National Natural Science Foundation of China (Contact No. 11775176), Basic Research Program of Natural Science of Shaanxi Province (Grant No. 2018KJXX-094), The Key Innovative Research Team of Quantum Many-Body Theory and Quantum Control in Shaanxi Province (Grant No. 2017KCT-12), and the Major Basic Research Program of Natural Science of Shaanxi Province (Grant No. 2017ZDJC-32).

\end{document}